\documentclass[12pt,a4paper,aps,amsmath,byrevtex,showpacs,showkeys]{revtex4}
\usepackage{bm}
\usepackage{graphics}
\usepackage{dcolumn}
\usepackage{enumerate}
\begin{document}

\title{New sum rules relating the 1-body momentum distribution of the 
homogeneous electron gas to the Overhauser 2-body wave functions of its 
pair density} 

\author{P. Ziesche}
\email[E-mail: ]{pz@mpipks-dresden.mpg.de}
\affiliation{Max-Planck-Institut f\"ur Physik komplexer Systeme\\
  N\"othnitzer Str. D-01187 Dresden, Germany}

\author{K. Pernal}
\email[E-mail: ]{pernalk@sus.univ.szczecin.pl}
\affiliation{Institute of Physics, University of Szczecin, Wielkopolska 15, 
70-451 Szczecin, Poland}

\author{F. Tasn\'adi}
\email[E-mail: ]{tasnadi@hp1.ifw-dresden.de}
\affiliation{Leibniz-Institut f\"ur Festk\"orper- und Werkstoffforschung 
Dresden, Germany \\
and University of Debrecen, Hungary}
\date{\today}

%*****************************************************************************************
  \begin{abstract}
The recently derived sum rules for the scattering phase shifts of the 
Overhauser
geminals (being 2-body-wave functions which parametrize the pair density  
together with an appropriately chosen occupancy) are generalized to integral
equations which allow in principle to calculate the momentum distribution, 
supposed the phase shifts of the  Overhauser geminals are known from an 
effective 
parity-dependent interaction potential (screened Coulomb repulsion).
\end{abstract}
%*****************************************************************************************

\maketitle

\newcommand{\integral}{\mathop{\int d^3r_2}\limits}
\newcommand{\integralp}{\mathop{\int d^3r'}\limits}

%*****************************************************************************************

\section*{Introduction}

The homogeneous electron gas is a model which allows to study pure electron
correlation without any interference with the multiple-scattering problem of
real molecules, clusters, and solids. \cite{Ful, Mar1} The 
quantum kinematics of this phenomenon is hidden in the reduced density matrices 
and their diagonals. \cite{Davi,Erd,Cio,Col} The simplest reduced densites are 
(only the spin-unpolarized ground state is considered here)
the 1-body momentum distribution $n(k)$, recently parametrized in terms of the 
convex Kulik function, \cite{Zie1,Gor1} and the 2-body quantities 
$g_{\uparrow\uparrow}(r)$ and $g_{\uparrow\downarrow}(r)$, being the 
non-negative pair densities (PDs) for electron pairs with parallel respectively
antiparallel spins and with an interelectronic distance $r$. 
\cite{Per,Gor2,Gor3} All these quantities depend parametrically on the electron
density $3/(4\pi r_s^3)$. Correlation induced properties of $n(k)$ (normalized 
as $(2/N)\sum_{\mathbf{k}}n(k)=1$) are: (i) its nonidempotency $0<n(k)<1$, 
measured by a quantity $c=(2/N)\sum_{\mathbf{k}}n(k)[1-n(k)]<1$ (which is 
christened here L\"owdin 
parameter \cite{Low}) and (ii) the quasi-particle weight $z_{\rm F}=
n(1^-)-n(1^+)<1$ ($k$ is measured in units of the Fermi wave length 
$1/(\alpha r_s), \alpha=(4/9\pi)^{1/3}$). Correlation induced properties of the
PDs with their asymptotics $g_{\uparrow\uparrow,\uparrow\downarrow}(\infty)=1$ 
are besides the oscillatory behavior for $r\to \infty$, for small $r$ the Fermi
hole $g_{\uparrow\uparrow}(r)<1$ with its on-top properties
$g_{\uparrow\uparrow}(0)=0$, $g'_{\uparrow\uparrow}(0)=0$, and
a characteristic curvature $g''_{\uparrow\downarrow}(0)>0$ and the 
Coulomb hole $g_{\uparrow\downarrow}(r)<1$ with its characteristic on-top value
$g_{\uparrow\uparrow}(0)<1$. The on-top curvature of the Fermi hole is a local 
measure of the correlation strength. \cite{Zie2,foo1,Dob,Luk,Fla} With the 
spin-summed PD $g=[g_{\uparrow\uparrow}+g_{\uparrow\downarrow}]/2$, 
particle-number fluctuations in spatial parts (fragments, domains) of the 
system can be discussed as another correlation index with the conclusion
"correlation suppresses fluctuations". \cite{Zie3,Zie4,Zie5,foo2,Sch,Sav,Poa} 
For the ideal Fermi gas ($r_s=0$) it is $n^{(0)}(k)=1-\Theta (k)$, $c^{(0)}=0$,
$z_{\rm F}^{(0)}=1$, and $g_{\uparrow\downarrow}^{(0)}(r)=1$. 

The virial theorem \cite{Mar2} provides a relation between the kinetic and the 
interaction energy, which follow from $n(k)$ and $g(r)$, respectively, giving
thus for their $r_s$ dependence an integral relation. \cite{Mac} In this paper
another relation between $n(k)$ and $g(r)$ is derived. This derivation is based
on the successful parametrization of the PDs in terms of Overhauser 2-body wave
functions (geminals), which are the scattering state solutions of an effective 
2-body Schr\"odinger equation with an appropriately screened Coulomb repulsion. 
\cite{Ove,Gor4,Gor5,Davo,Zie6} It is furthermore based on the assumption that
these Overhauser geminals can be used to represent also the 2-body reduced 
density matrix (2-matrix) $\gamma_2(1|1',2|2')$, the digonal of which gives the
PD. Now the idea was to obtain the 1-matrix $\gamma_1$ from the contraction 
of the 2-matrix $\gamma_2$. Unfortunately for extended systems this does not 
work, because
this contraction is not size-extensive. This problem is easy to overcome with 
the help of the cumulant expansion. \cite{Zie7,Zie5,foo3,Zie8,Zie9} It  
defines by $\gamma_2=A\gamma_1\gamma_1-\chi$ ($A=$ antisymmetrizer) the 
size-extensively normalizable and contractable cumulant 2-matrix $\chi =
A\gamma_1\gamma_1-\gamma_2$, which is here thus represented in terms of Bessel 
functions (from $A\gamma_1\gamma_1$) and Overhauser geminals (from $\gamma_2$).
From the normalization of $\chi$ follow sum rules (SRs) for the scattering 
phase shifts of the Overhauser geminals. \cite{Zie6} These (Friedel like) SRs 
are generalized in this paper by using the contraction properties of $\chi$, 
such that (at least in principle) $n(k)$ can be calculated supposed the 
phase shifts of the Overhauser geminals are known.

\section*{Basic notation and normalization sum rules}
The singlet($+$ for even $l$)/triplet ($-$ for odd $l$) components of the PD in terms of Overhauser geminals
$R_l(r,k)$ are
\begin{equation}
g_{\pm}(r)=\sum\nolimits_L^{\pm} < \mu(k)R_l^2(r,k) >, \qquad L=(l,m_l).
\label{OHPD}
\end{equation}
The $k$-average is defined by 
\begin{equation}
< \dots > = \frac{2}{N}\sum\nolimits_{\mathbf{k}} \dots =
\frac{2}{N} \int_0^\infty\frac{\Omega d^3k}{(2\pi)^3} \dots =
\int_0^{\infty} d(k^3) \cdots 
\label{Ksum}
\end{equation}
with the normalization volume $\Omega$ and the density $\varrho=N/\Omega=
1/3\pi^2$. Wave lengths are measured in units of the Fermi wave length 
$k_{\rm F}=1/(\alpha r_s)$, $\alpha= (4/(9\pi))^{1/3}$. In this paper the 
Overhauser occupancy is slightly generalized by
\begin{equation}
\mu(k)=\frac{2}{N} \sum\nolimits_{\mathbf{K}} \mu(\mathbf{K},\mathbf{k}), \qquad
\mu(\mathbf{K},\mathbf{k})=n(\Big| \frac{1}{2}\mathbf{K}+\mathbf{k} 
\Big|)n(\Big| \frac{1}{2}\mathbf{K}-\mathbf{k} \Big|),
\label{Oocc}
\end{equation}
because it arises here not from the idempotent 
$n^{(0)}(k)=\Theta(1-k)$ of the ideal Fermi gas, but from the nonidempotent 
$n(k)$ of the interacting electron gas. \cite{Zie1,Gor1}
It is $\mu(0)=8 (1-c)$. The $R_L(\mathbf{r},k)=
R_l(r,k)Y_L(\mathbf{e}_r)$ are the (scattering state) solutions of
the 2-body Schr\"odinger equation (the center-of-mass motion separates 
completely)
\begin{equation}
\lbrack -\Delta + v_{\pm}(r) -k^2 \rbrack R_L(\mathbf{r},k)=0, \qquad v_{\pm}(\infty)=0
\label{SEQ}
\end{equation}
with an appropriate repulsive interaction potential $v_{\pm}(r)=
\frac{\alpha r_s}{r}+\cdots$, possibly different for \\
$"+"$ or $"-"$. \cite{Ove,Gor4,Gor5,Davo,Zie6} The success of the
Overhauser approach means, that local interaction potentials $v_+(r)$ and
$v_-(r)$, if not being exact, are at least reasonable approximations. 

One generalization of Eq.(\ref{OHPD}) concerns inhomogeneous systems. 
\cite{Davo} Here another generalization is considered, namely the 
representation of the 2-matrix in terms of Overhauser geminals:
\begin{equation}
\gamma_{\pm}(\mathbf{R} \vert \mathbf{R}',\mathbf{r} \vert \mathbf{r}')=
\varrho^2(4\pi)^2 \sum\nolimits_{L,L'}^{\pm} 
< 
\tilde{\mu}_{LL'}(\mathbf{R}-\mathbf{R}',k) R^{}_L(\mathbf{r},k) R^{\ast}_{L'}(\mathbf{r}',k) 
>
\label{2matrix}
\end{equation}
$\mathbf{R}=\frac{1}{2}(\mathbf{r}_1+\mathbf{r}_2)$ is the 
center-of-mass coordinate and $\mathbf{r}=\mathbf{r}_1-\mathbf{r}_2$ is the 
relative coordinate. As $\mu(k)$, the $l$-independent but $k$-dependent weight 
in Eq.(\ref{OHPD}), also here the occupancy matrix
\begin{eqnarray}
\tilde{\mu}_{LL'}(\mathbf{R},k)=
\frac{2}{N}\sum\nolimits_{\mathbf{K}} 
{\rm e}^{{\rm i}\mathbf{K}\mathbf{R}}\mu_{LL'}(\mathbf{K},k) , \quad 
%\nonumber \\
\mu_{LL'}(\mathbf{K},k)=
\int\frac{d\Omega_k}{4\pi} \ Y^{\ast}_{L}(\mathbf{e}_k) \mu(\mathbf{K},\mathbf{k})
Y_{L'}(\mathbf{e}_k)
\label{OccMatrix}
\end{eqnarray}
follows with Eq. (\ref{Oocc}) from the momentum distribution $n(k)$.

Arguments in favour of Eq.(\ref{2matrix}) are:
(i) The diagonal elements give the Overhauser PD Eq.(\ref{OHPD}), 
$\gamma(\mathbf{R} \vert \mathbf{R},\mathbf{r} \vert \mathbf{r})=
\varrho^2 g_{\pm}(r)$, and 
(ii) in the cumulant partitioning of $\gamma_{\pm}$ the generalized HF term 
$\gamma_{\pm}^{\rm HF}$
is given by the same expression as Eq.(\ref{2matrix}) with only $R_L(\mathbf{r},k)$ replaced
by $j_L(k\mathbf{r})=j_l(kr)Y_L(\mathbf{e}_r)$ because the natural orbital are 
plane waves and using the $L$-expansion of a plane wave, 
${\rm exp}({\rm i}\mathbf{k}\mathbf{r})= 4\pi\sum\nolimits_L {\rm i}^l 
j_L(k\mathbf{r})Y_L^*(\mathbf{e}_k)$.

So the cumulant 2-matrix $\chi_{\pm}=-(\gamma_{\pm}-\gamma_{\pm}^{\rm HF})$ is given by
\begin{equation}
\chi_{\pm}(\mathbf{R} \vert \mathbf{R}',\mathbf{r} \vert \mathbf{r}')=
-\varrho^2(4\pi)^2 \sum\nolimits_{L,L'}^{\pm} 
<
\tilde{\mu}_{LL'}(\mathbf{R}-\mathbf{R}',k) \lbrack R^{}_L(\mathbf{r},k) R^{\ast}_{L'}(\mathbf{r}',k)
- j^{}_L(k\mathbf{r}) j^{\ast}_{L'}(k\mathbf{r}') \rbrack
>
\label{Cum2matrix}
\end{equation}

For the normalization 
\begin{eqnarray}
\int d^3r \ \chi_{\pm}(\mathbf{R} \vert \mathbf{R},\mathbf{r} \vert \mathbf{r})
& = & -\varrho^2 \int d^3r \  \sum\nolimits_{L}^{\pm} 
< \mu(k) \lbrack R_l^2(r,k)- j_l^2(kr) \rbrack  > \nonumber \\
& = &\varrho \frac{2}{\pi}\sum\nolimits_{L}^{\pm} \int_0^{\infty} dk \ \mu(k) \eta_l'(k)
=\pm \varrho c ,  
\label{normalization}
\end{eqnarray}
%where the nonidempotency measure 
%\begin{equation}
%c=\frac{2}{N}\sum\nolimits_{\mathbf{k}} n(k) \lbrack 1-n(k) \rbrack
%\end{equation}
%is christened L\"owdin parameter.\cite{Low} 
cf. Ref. \cite{Zie6}.

\section*{Contraction sum rules}
With the contracted cumulant 
2-matrices and their Fourier transform
\begin{equation}
\chi_{\pm}(\vert \mathbf{r}_1 - \mathbf{r}'_1 \vert)=\integral_{r^{}_2<R \to \infty}
\chi_{\pm}(\mathbf{R} \vert \mathbf{R}',\mathbf{r} \vert \mathbf{r}')
\Big|_{\mathbf{r}'_2 = \mathbf{r}^{}_2}, \quad
\tilde{\chi}_{\pm}(\kappa)=\frac{1}{2}\varrho \int d^3r \  
{\rm e}^{-{\rm i}\boldsymbol{\kappa}\mathbf{r}}{\chi}_{\pm}(r),
\label{ContCum2matrix}
\end{equation}
the contraction SRs are  
\begin{equation}
\tilde{\chi}_{\pm}(\kappa) = 
\mp \ \varrho \ n(\kappa) \lbrack 1-n(\kappa)\rbrack. 
\label{SR}
\end{equation}
They follow from ${\tilde \gamma}_{\pm}(\kappa)=\varrho\ n(\kappa)
[\frac{1}{2}N\pm 1]$, ${\tilde \gamma}_{\pm}^{\rm HF}(\kappa)=
\varrho\ n(\kappa) [\frac{1}{2}N\pm n(\kappa)]$, and ${\tilde \chi}_\pm=
{\tilde \gamma}_\pm^{\rm HF}-{\tilde \gamma}_\pm$, showing explicitly the 
cancellation of the non-size extensive terms. Eq. (\ref{SR}) says that $n(k)$ 
can be calculated, if the lhs is known. With 
$\chi_{\pm}(0)=\frac{2}{N}\sum_{\boldsymbol{ \kappa}}{\tilde \chi}_{\pm}
(\kappa)=\mp \varrho c$ the special phase shift SRs of Ref. \cite{Zie6} are 
contained in the more general SRs (\ref{SR}) as special cases.

With the identity (following from Eq. (\ref{SEQ}) and generalizing Eq. (13) of 
Ref. \cite{Zie6})
\begin{eqnarray}
R^{}_LR'^{\ast}_{L'}=
\left( \frac{\partial }{\partial \mathbf{r}}+\frac{\partial}{\partial \mathbf{r}'} \right)
\frac{1}{2}\left[  \frac{\partial R^{}_L}{\partial \mathbf{r}} \stackrel{\bullet}{R'^{\ast}_{L'}}- 
R^{}_L \frac{\partial \stackrel{\bullet}{R'^{\ast}_{L'}}}{\partial \mathbf{r}'}+{\rm h.c.}\right] \nonumber \\
-\frac{1}{2}[ v_{\pm}(r)-v_{\pm}(r')] 
\left[ R^{}_L \stackrel{\bullet}{R'^{\ast}_{L'}}-\stackrel{\bullet}{R^{}_L} R'^{\ast}_{L'} \right],
\label{identity}
\end{eqnarray}
(where $R'_{L'}=R_{L'}(\mathbf{r}',k)$, $\stackrel{\bullet}{R_L}=
\partial R_L(\mathbf{r},k)/\partial k^2$, and h.c. means complex conjugate 
together with an exchange in both coordinates and indices) from 
Eqs.(\ref{normalization}) and (\ref{ContCum2matrix}) it turns out 
\begin{equation}
\chi_{\pm}(r)=\chi_{\pm}^A(r)+\chi_{\pm}^B(r)=-\varrho^24\pi \sum\nolimits_{L,L'}^{\pm}
<
\tilde{\mu}_{LL'}\left(\frac{1}{2}\mathbf{r},k \right) \left[ A_{LL'}(\mathbf{r},k)+B_{LL'}
(\mathbf{r},k)\right]
>.
\end{equation}
The $A$ and $B$ matrices are defined by
\begin{equation}
A_{LL'}(\mathbf{r},k)+B_{LL'}(\mathbf{r},k)=
4\pi \integralp_{r'<R\to \infty}
\left[ R^{}_L(\mathbf{r}+\mathbf{r}',k) R^{\ast}_{L'}(\mathbf{r}',k)-
j^{}_L(k(\mathbf{r}+\mathbf{r}')) j^{\ast}_{L'}(k\mathbf{r}')\right]
\label{ABmatrix}
\end{equation}
where $A_{LL'}$ results from the 1st term of Eq.(\ref{identity}) and $B_{LL'}$ 
from the 2nd one. With the Gauss theorem it is 
\begin{equation}
A_{LL'}(\mathbf{r},k)=\delta_{LL'}4\pi R^2 
\left[ \frac{\partial R_l}{\partial R}\stackrel{\bullet}{R_l}-R_l\frac{\partial}{\partial R}\stackrel{\bullet}{R_l}-
\frac{\partial j_l}{\partial R}\stackrel{\bullet}{j_l}+j_l\frac{\partial}{\partial R}\stackrel{\bullet}{j_l}
 \right]_{R \to \infty},
\label{Aterm}
\end{equation}
(where $R_l=R_l(R,k)$ and the $\mathbf{r}$-dependence disappears) and with the 
analysis of Ref. \cite{Zie6} it is finally
\begin{equation}
A_{LL'}(\mathbf{r},k)=\delta_{LL'}A_l(k), \qquad A_l(k)=\frac{2\pi}{k^2}\eta_l'(k).
\label{FinalAterm}
\end{equation}
So
\begin{equation}
\chi_{\pm}^A(r)=-\varrho^2 \sum\nolimits_L^{\pm} \frac{2}{N}\sum_\mathbf{K}
{\rm e}^{{\rm i}\mathbf{K}\mathbf{r}/2}<\mu (K,k)A_l(k)> , \quad
\mu(K,k)=\int \frac{d\Omega_k}{4\pi}\mu (\mathbf{K},\mathbf{k}).
\end{equation}
Because of $\chi_{\pm}^{B}(0)=0$ and Eq. (\ref{Oocc}), it results
$\chi_{\pm}(0)=-\varrho^2\sum_L^{\pm}<\mu(k)A_l(k)>$, in agreement with the 
(Friedel like) phase shift SRs of Ref. \cite{Zie6}, saying 
$\chi_{\pm}(0)=\mp \varrho c$. The Fourier transform 
\begin{equation} 
\tilde{\chi}_{\pm}^A(\kappa)= 
-\varrho^2\sum\nolimits_L^{\pm} <2^3\mu(2\kappa,k)A_l(k)> , 
\label{FinalA}
\end{equation}
enters (together with the $B$ term) the more general SR (\ref{SR}).
 
As already mentioned, the $B$ matrix of Eq. (\ref{ABmatrix}) follows from the 
2nd term of Eq. (\ref{identity})
\begin{eqnarray}
B_{LL'}(\mathbf{r},k)=-4\pi \int d^3r'\  \frac{1}{2}
\left[ 
v_{\pm}(|\mathbf{r}+\mathbf{r}'|)-v_{\pm}(r')
\right] \times \nonumber \\
\left[
R^{}_L(\mathbf{r}+\mathbf{r}',k)\stackrel{\bullet}{R^{\ast}_{L'}}(\mathbf{r}',k)-
\stackrel{\bullet}{R^{}_L}(\mathbf{r}+\mathbf{r}',k) R^{\ast}_{L'}(\mathbf{r}',k)
\right].
\label{Bterm}
\end{eqnarray}
So
\begin{equation}
\chi_{\pm}^B(r)=-\varrho^2 4\pi \sum\nolimits_{L,L'}^{\pm} 
<\tilde{\mu}_{LL'}\left(\frac{1}{2}\mathbf{r},k \right)B_{LL'}(\mathbf{r},k)>
\end{equation}
and
\begin{equation}
\tilde{\chi}_{\pm}^B(\kappa)=-\frac{1}{2}\varrho^2 4\pi \sum\nolimits_{L,L'}^{\pm}
<\frac{2}{N}\sum\nolimits_{\boldsymbol{\kappa}_1}\mu_{LL'}(2({\boldsymbol{\kappa}}-
{\boldsymbol{\kappa}_1}),k)\tilde{B}_{LL'}({\boldsymbol{\kappa}_1},k)>.
\label{FTchiB}
\end{equation}
Note  $\sum\nolimits_{\boldsymbol{\kappa}}\tilde{\chi}^B_{\pm}(\kappa)=0$.
Thus the $B$ term does not contribute to the normalization of $n(k)$ as it does
not contribute to the normalization of the PDs.

The Fourier transformed interaction potential and Overhauser geminals
\begin{equation}
\tilde{v}_{\pm}(\kappa)=\varrho \int d^3r \ 
{\rm e}^{-{\rm i}\boldsymbol{\kappa}\mathbf{r}}v_{\pm}(r), \quad 
\tilde{R}_{L}(\boldsymbol{\kappa},k)
=\varrho \int d^3r \ {\rm e}^{-{\rm i}\boldsymbol{\kappa}\mathbf{r}}R_{L}(\mathbf{r},k)
\label{FTOG}
\end{equation}
determine the Fourier transformed $B$ matrix
\begin{equation}
\tilde{B}_{LL'}(\boldsymbol{\kappa}_1,k)= - \frac{4\pi}{4\varrho} \frac{2}{N} \sum\nolimits_{\boldsymbol{\kappa}_2} 
\tilde{v}_{\pm}(\kappa_{12})
\left[
\tilde{R}^{}_L(\boldsymbol{\kappa}_2,k)\stackrel{\bullet}{\tilde{R}^{\ast}_{L'}}(\boldsymbol{\kappa}_1,k)
-\stackrel{\bullet}{\tilde{R}^{}_L}(\boldsymbol{\kappa}_1,k)\tilde{R}^{\ast}_{L'}(\boldsymbol{\kappa}_2,k)
\right], 
\label{FTBmatrix}
\end{equation}
$\kappa_{12}=|\boldsymbol{\kappa}_1-\boldsymbol{\kappa}_2|$.
Eq.(\ref{FTBmatrix}) has to be inserted into Eq.(\ref{FTchiB}). Then with 
($\zeta_{12}=\mathbf{e}_1\mathbf{e}_2$) 
\begin{equation}
v_{\pm}(\kappa_{12})=\sum\nolimits_{L''}P_{l''}(\zeta_{12})
v_{l''}^{\pm}(\kappa_1,\kappa_2)
\quad \text{or} \quad
v_{l''}^{\pm}(\kappa_1,\kappa_2)=\int_{-1}^{1} \frac{d \zeta_{12}}{2} \ P_{l''}(\zeta_{12})v_{\pm}(\kappa_{12})
\end{equation}
and with 
\begin{equation}
\int \frac{d\Omega_\kappa}{4\pi} \ \mu(2(\boldsymbol{\kappa}-\boldsymbol{\kappa}_1),\mathbf{k})=
\sum\nolimits_{L_1} P_{l_1}(\mathbf{e}_1\mathbf{e}_k)\mu_{l_1}(\kappa,\kappa_1,k)
\end{equation}
(the invariance of $\mu(\mathbf{K},\mathbf{k})$ by the replacement 
$\mathbf{k}\rightarrow -\mathbf{k}$ makes $l_1$ even) and with 
\begin{equation}
%(4\pi)^2 \ \sum\nolimits_{L,L'}^{\pm} \sum\nolimits_{L_1}^{}
%\left|C_{L L_1 L'}\right|^2\dots =\sum\nolimits_{L,L_1,L'}^{\pm} C_{l l_1 l'} 
%\dots, \quad
C_{ll_1l'}=\int d\Omega \ P_l(\zeta)P_{l_1}(\zeta)P_{l'}(\zeta) 
\end{equation}
(because $l$ and $l'$ have the same parity, these coefficients vanish for odd
$l_1$) an expression arises, 
%the final form of the terms entering the contraction SRs (\ref{SR}) is 
%Eq. (\ref{FinalA}) for the $A$ term and 
\begin{eqnarray}
\tilde{\chi}_{\pm}^{B}(\kappa) & = & \varrho
\sum\nolimits_{L_,L'}^\pm \sum\nolimits_{L_1}C_{l l_1 l'} \;
\left (\frac{2}{N}\right )^2\sum\nolimits_{\boldsymbol{\kappa}^{}_{1,2}} 
<\mu_{l_1}(\kappa,\kappa_1,k) \times \nonumber \\
& & \left[ v_l^{\pm}(\kappa_1,\kappa_2)\tilde{R}_l(\kappa_2,k)
\stackrel{\bullet}{\tilde{R}_{l'}}(\kappa_1,k)
-\stackrel{\bullet}{\tilde{R}_l}(\kappa_1,k)
\tilde{R}_{l'}(\kappa_2,k) v_{l'}^{\pm}(\kappa_1,\kappa_2)\right] > , 
\label{FinalB}
\end{eqnarray}
which vanishes, because the $L$-$L'$ double sum runs over an antisymmetric
matrix (each element $l,l'$ is compensated by its mirror party). Thus, it 
finally results   
\begin{equation}
\frac{2}{\pi}\sum\nolimits_L^{\pm} <\frac{2^3\mu(2\kappa,k)}{3k^2}\eta '_l(k)>
=\pm n(k)[1-n(k)].
\label{Final}
\end{equation}
The occupancy weight in Eq. (\ref{Final}) can 
be written in terms of $n(k)$ as
\begin{eqnarray}
\mu(2\kappa,k) &=& 
\int\frac{d\Omega_k}{4\pi}
n(|\boldsymbol{\kappa}+\mathbf{k}|) n(|\boldsymbol{\kappa}-\mathbf{k}|), 
%\mu_l(\kappa,\kappa_1,k) &=& 
%\int\frac{d\Omega_1}{4\pi}
%\int\frac{d\Omega_k}{4\pi}P_l(\mathbf{e}_1\mathbf{e}_k)
%n(|\boldsymbol{\kappa}-\boldsymbol{\kappa}_1+\mathbf{k}|)
%$n(|\boldsymbol{\kappa}-\boldsymbol{\kappa}_1-\mathbf{k}|).
\end{eqnarray}
or with $n(k)=\frac{1}{2}\rho\int d^3r\;
\frac{\sin \kappa r}{\kappa r}f(r)$ in terms of the dimensionless 1-matrix 
$f(r)$ as
\begin{eqnarray}
\mu(2\kappa,k) 
& = & \frac{1}{4}\rho^2\int d^3r_1d^3r_2\; 
\frac{\sin \kappa |\mathbf{r}_1+\mathbf{r}_2|}
{\kappa |\mathbf{r}_1+\mathbf{r}_2|}
\frac{\sin k|\mathbf{r}_1-\mathbf{r}_2|}
{k|\mathbf{r}_1-\mathbf{r}_2|}f(r_1)f(r_2),  
%\mu_l(\kappa,\kappa_1,k) 
%& = & \frac{1}{4}\rho^2\int d^3r_1d^3r_2\; 
%P_l\left (\frac{r_1^2-r_2^2}
%{|\mathbf{r}_1+\mathbf{r}_2||\mathbf{r}_1-\mathbf{r}_2|}\right )\times
%\nonumber \\
%& & \frac{\sin \kappa|\mathbf{r}_1+\mathbf{r}_2|}
%{\kappa |\mathbf{r}_1+\mathbf{r}_2|}
%j_l(\kappa_1|\mathbf{r}_1+\mathbf{r}_2|)
%j_l(k|\mathbf{r}_1-\mathbf{r}_2|)f(r_1)f(r_2).
\end{eqnarray}
For the momentum distribution the parametrization in terms of the Kulik function
can be used. \cite{Zie1,Gor1}

\section*{Conclusions}
The contraction SRs (\ref{Final}) 
are nonlinear integral equations for $n(k)$ supposed the phase shifts of the
Overhauser geminals  $R_l(r,k)$ are known from the effective
interaction potential $v_{\pm}(r)=\frac{\alpha r_s}{r}+\cdots$. These integral
equations have to be solved selfconsistently. Because the PD is (via the
$R_l(r,k)$) a functional of $v_{\pm}(r)$ and $n(k)$ is a functional of the
$R_l(r,k)$, thus the total energy becomes a functional of $v_{\pm}(r)$. \\ 

The success of the Overhauser approach seems to confirm the
possibility of a pair-density functional theory, the idea of which is presented
in Ref. \cite{Zie10}. Whereas therein the geminal occupancy was erroneously 
assumed to be 1 and 0 according to occupied and unoccupied geminals 
respectively (analog to the aufbau principle of the Hartree-Fock and the 
Kohn-Sham schemes), in the Overhauser approach a geminal occupancy $\mu(k)$ is 
used being quite different from a step function. For PD or 2-matrix 
geminals there is no aufbau principle. 
%******************************************************************************************

\section*{Acknowledgments}
One of the authors (P.Z.) gratefully acknowledges J.P. Perdew, P. Gori-Giorgi, 
and M.P. Tosi for inspiring discussions and P. Fulde for supporting this work.
Another author (K.P.) expresses her thanks to the Max Planck Institute for the
Physics of Complex Systems for hospitality and F.T. thanks 
H. Eschrig for his support of this work and Technische Universit\"at Dresden 
for a scholarship.


\begin{thebibliography}{99}
\frenchspacing


\bibitem{Ful}     P. Fulde, {\it Electron Correlation in Molecules and
                  Solids}, 3rd ed., Springer, Berlin, 1995.

\bibitem{Mar1}    N.H. March, {\it Electron Correlation in Molecules and 
                  Condensed Phases}, Plenum, New York, 1996; N.H. March (ed.),
                  {\it Electron Correlation in the Solid State}, Imperial
                  College Press, London, 1999.


\bibitem{Davi}    E.R. Davidson, {\it Reduced Density Matrices in Quantum 
                  Chemistry}, Academic Press, New York, 1976.

\bibitem{Erd}     R. Erdahl and V.H. Smith, Jr., {\it Density Matrices and 
                  Density Functionals}, Reidel, Dordrecht, 1987.

\bibitem{Cio}     J. Cioslowski (ed.), {\it Many-Electron Densities and 
                  Reduced Density Matrices}, Kluwer/Plenum, New York, 2000.

\bibitem{Col}     A. J. Coleman and V.I. Yukalov, {\it Reduced Density 
                  Matrices}, Springer, Berlin, 2000. 


\bibitem{Zie1}   P. Ziesche, phys. stat. sol. (b) {\bf 232}, 231 (2002).

\bibitem{Gor1}   P. Gori-Giorgi and P. Ziesche, Phys. Rev. B, Dec 15, 2002, in
                  press.


\bibitem{Per}   J. P. Perdew and Y. Wang, Phys. Rev. B {\bf 46}, 12947
                (1992); {\bf 56}, 7018 (1997).

\bibitem{Gor2}  P. Gori-Giorgi, F. Sacchetti, and G.B. Bachelet,
                Phys. Rev. B {\bf 61}, 7353 (2000), B {\bf 66}, 159901(E)

\bibitem{Gor3}  P. Gori-Giorgi and J. P. Perdew, Phys.
                Rev. B {\bf 66}, 165118 (2002).


\bibitem{Low}   P.-O. L\"owdin, Adv. Chem. Phys. {\bf 2}, 207 (1959).

\bibitem{Zie2}  P. Ziesche, J. Mol. Struc. (Theochem) {\bf 527}, 35 (2000),
                  therein factors 1/2 are erroneously incorporated in the
                  definition of $g_{\uparrow\uparrow}(r)$
                  and $g_{\uparrow\downarrow}(r)$, such that
                  $g_{\uparrow\uparrow}(\infty)=
                  g_{\uparrow\downarrow}(\infty) = 1/2$ instead of 1.


\bibitem{foo1}  For inhomogeneous (in particular finite) systems the on-top 
                curvature of the Fermi hole has 
                been discussed a lot in terms of electron localization, of its 
                relation to chemical shell structure and bonding, and of Fermi 
                hole mobility by Daudel, Bader, Stoll, Colle and Salvetti, 
                Luken, Becke, Dobson, Savin, and others, cf. e.g.
                \cite{Dob,Luk,Fla}, further refs. are in Ref. \cite{Zie2}.

\bibitem{Dob}   J.F. Dobson, J. Chem. Phys. {\bf 94}, 4328 (1991) and refs.
                therein. 

\bibitem{Luk}   W.L. Luken, in {\it Theoretical Models of Chemical Bonding},
                part 2, Z. Maksic (ed.), Springer, Berlin, 1999 and refs. 
                therein.

\bibitem{Fla}   H.-J. Flad, F. Schautz, Y. Wang, M. Dolg, and A. Savin, 
                Eur. Phys. J. D{\bf 6}, 243 (1999) and refs. therein.

\bibitem{Zie3}  P. Ziesche, in {\it Electron Correlations and Materials
                Properties}, edited by A. Gonis, N. Kioussis, and
                M. Ciftan, Kluwer/Plenum, New York, 1999, p. 361.

\bibitem{Zie4}  P. Ziesche, J. Tao, M. Seidl, and J. P. Perdew,
                Int. J. Quantum Chem. {\bf 77}, 819 (2000).

\bibitem{Zie5}   P. Ziesche, in \cite{Cio}, p. 33.

\bibitem{foo2}   For finite systems atomic charge fluctuations have been 
                discussed by Fulde el al., Savin et al., Fradera et al. as 
                correlation measures \cite{Ful}, p. 157,
                and \cite{Sch,Sav} or as delocalization index. \cite{Poa} 

\bibitem{Sch}   F. Schautz, H.-J. Flad, and M. Dolg, Theor. Chem. Acc. {\bf 99},
                231 (1998).

\bibitem{Sav}   A. Savin, in {\it
                Review in Modern Quantum Chemistry}, K. D. Sen (ed.), World
                Scientific, Singapore, 2002, and refs. therein.

\bibitem{Poa}   J. Poater, X. Fradera, M. Duran, and M. Sol$\grave {\rm a}$, 
                Chem. Eur. J. {\bf 8}, in press, and refs. therein.

\bibitem{Mar2}  N.H. March, Phys. Rev. {\bf 110}, 604 (1958). 

\bibitem{Mac}   W. Macke and P. Ziesche, Ann. Physik (Leipzig) {\bf 13}, 25
                (1964). 






\bibitem{Ove}  A.~W.~Overhauser, Can. J. Phys. {\bf 73}, 683 (1995).

\bibitem{Gor4}  P. Gori-Giorgi and J. P. Perdew, Phys. Rev. B
                {\bf 64}, 155102 (2001).

\bibitem{Gor5}  P. Gori-Giorgi, cond-mat/0111141 and in {\it Electron 
                Correlations and Materials
                Properties} II, edited by A. Gonis, N. Kioussis, and
                M. Ciftan, Kluwer/Plenum, New York, in press.

\bibitem{Davo}   B. Davoudi, M. Polini, R. Asgari, and M.P. Tosi,
                cond-mat/0201423, and Phys. Rev. B {\bf 66}, 075110 (2002). In
                a subsequent
                paper (cond-mat/0206456) the same authors go beyond the
                Hartree-like theory and show how the inclusion of exchange
                and correlation through a Kohn-Sham equation let
                emerge liquid-like structures with increasing coupling
                strength $r_s$ through the formation of a first-neighbor shell
                and further oscillations in the PD $g(r)$, which for
                larger $r_s$ becomes a precursor of the Wigner 
                crystallisation.

\bibitem{Zie6}   P. Ziesche, cond-mat/0211408.


\bibitem{Zie7} P. Ziesche, Solid State Commun. {\bf 82}, 597 (1992);

\bibitem{foo3}   Another advantage of the cumulant expansion is that $\chi$ is
                 given perturbatively by (the also size-extensive) linked 
                 diagrams. In this way $\chi$ has been studied recently in its
                 high-density limit. \cite{Zie8,Zie9}  

\bibitem{Zie8}   P. Ziesche, Int. J. Quantum Chem. {\bf 90}, 342 (2002).

\bibitem{Zie9}   P. Ziesche, in {\it Electron Correlations and Materials
                Properties} II, edited by A. Gonis, N. Kioussis, and
                M. Ciftan, Kluwer/Plenum, New York, in press. 

\bibitem{Zie10} M. Levy and P. Ziesche, J. Chem. Phys. {\bf 115}, 9110 (2001)
                 and refs. therein. 

%\bibitem{Kim}   J. C. Kimball, J. Phys. a {\bf 8}, 1513 (1975).

%\bibitem{Yas}   H. Yasuhara and Y. Kawazoe, Physica A {\bf 85}, 416 (1976).


\end{thebibliography}
\end{document}